\documentstyle[preprint,epsfig,aps]{revtex}

\begin{document} 

\title{A large Hilbert space QRPA and RQRPA calculation \\
of neutrinoless double beta decay
\footnote{Supported by the ``Deutsche 
Forschungsgemeinschaft'', contract No. Fa 67/17-1 and
by the ``Graduierten Kolleg - Struktur und Wechselwirkung 
von Hadronen und Kernen'', DFG, Mu 705/3.}}

\author{F. \v Simkovic$^1$
\thanks{{\it On leave from:}Bogoliubov Theoretical Laboratory, 
Joint Institute for Nuclear Research, 
141980 Dubna, Moscow Region, Russia and Department of Nuclear Physics,  
Comenius University, Mlynsk\'a dolina F1, Bratislava, Slovakia}, 
J. Schwieger$^1$, 
G. Pantis$^2$ and Amand Faessler$^1$ }
\address{1.  Institute f\"ur Theoretische Physik der Universit\"at 
T\"ubingen\\ 
Auf der Morgenstelle 14, D-72076 T\"ubingen, Germany \\
2.  Theoretical Physics Section, University of Ioannina,\\
GR 451 10, Ioannina, Greece }
\date{\today}
\maketitle
\begin{abstract}
A large Hilbert space is used for the calculation of  
the nuclear matrix elements
governing the light neutrino mass mediated mode of
neutrinoless double beta decay ($0\nu\beta\beta$-decay) of $^{76}Ge$,
$^{100}Mo$, $^{116}Cd$, $^{128}Te$ and $^{136}Xe$ within the 
proton-neutron  quasiparticle random phase approximation (pn-QRPA) and 
the renormalized QRPA with proton-neutron pairing (full-RQRPA) methods.
We have found that the nuclear matrix elements  obtained with the 
standard pn-QRPA for several nuclear transitions are extremely 
sensitive to the renormalization of the particle-particle component
of the residual interaction of the nuclear hamiltonian. Therefore the
standard pn-QRPA does not guarantee the necessary accuracy to  allow
us to extract a reliable limit on the effective neutrino mass.  
This behaviour, already known  from the calculation of the two-neutrino
double beta decay matrix elements, 
manifests itself in the neutrinoless double-beta decay but
only if a large model space is used.
The full-RQRPA, which takes into account proton-neutron pairing and 
considers the Pauli principle in an approximate way, offers a stable 
solution in the physically acceptable region of the particle-particle
strength. In this way  more accurate values on the effective neutrino
mass have been deduced from the experimental lower limits of the
half-lifes of neutrinoless double beta decay. 
\end{abstract}

\section{Introduction}

The neutrinoless double beta decay ($0\nu\beta\beta$-decay) continues to
attract the attention of both experimentalists  and theoreticians
for a long period. In this process the nucleus (A,Z) undergoes
the transition to nucleus (A,Z+2) with the emission of two electrons:
\begin{equation}
(A,Z) ~ \rightarrow ~ (A,Z+2) ~ + 2e^-.
\end{equation}
It is obvious that this process violates the lepton number $L_e$ by two
units and is forbidden in the Standard Model. The background considerations 
imply that the $0\nu\beta\beta$-decay is measured on nuclei for which 
the ordinary single beta decay is either forbidden by energy law conservation 
or strongly inhibited by large spin changes. The $0\nu\beta\beta$-decay 
has not been seen in the experiment till now. A possible detection would be  
undoubtedly a signal of new physics beyond the standard model.  
The experimental lower limits 
provide us e.g.  with the most stringent limits on the effective 
neutrino mass, and the parameters of right-handed currents and coupling
constants of the supersymetric particles. If neutrinos turn out to be 
massive, the $0\nu\beta\beta$-decay experiment is considered to be the most 
sensitive to the existence of Majorana neutrinos coupled to electron.
It is wortwhile to notice that there is another two-neutrino mode of 
double beta ($2\nu\beta\beta$-decay) with two antineutrinos and electrons
in the final state, which is allowed within the Standard Model. This mode
being independent of the unknown particle physics parameters serves as a 
sensitive test of nuclear structure calculations.
There exist extensive reviews of the theory and phenomenology of the double
beta decay and we refer the  interested reader e.g., to Refs. \cite{1}-\cite{5}
for details. New 
contributions to the R-parity violating ($R_p \hspace{-1em}/\;\:$) 
supersymmetry (SUSY) mechanisms in $0\nu\beta\beta$-decay are  discussed
in Ref. \cite{6}.

In order to deduce the limits on the parameters from the particle physics point
of view,
it is necessary to calculate the corresponding nuclear matrix elements. 
The proton-neutron Quasiparticle Random Phase
Approximation (pn-QRPA) has been considered the most practical 
method for nuclear structure calculations of nuclear systems 
which are far away from closed shells \cite{7}-\cite{14}. 
However, the extreme sensitivity of the calculated $2\nu\beta\beta$-decay 
matrix elements in the physically acceptable region of the 
particle-particle strength of the nuclear Hamiltonian 
renters it difficult to  make definite rate predictions \cite{12}-\cite{14}.
This quenching behavior of the $2\nu\beta\beta$-decay matrix elements is a
puzzle and has attracted the attention of many theoreticians.
Other shortcomings of the pn-QRPA 
(e.g.,particle number non-conservation, the question of the 
proton-neutron pairing,
the violation of the Pauli exclusion principle etc. ) indicate that 
we have to go beyond the pn-QRPA in the evaluation of the 
double beta decay nuclear matrix elements.
Toivanen and Suhonen have proposed a proton-neutron
renormalized QRPA (pn-RQRPA) to study the double beta decay \cite{15}.
The pn-QRPA
is based on the renormalized quasiboson approximation, which considers 
the Pauli exclusion principle in an approximate way. 
Schwieger, \v Simkovic and  Faessler have extended the pn-RQRPA
to include proton-neutron pairing (full-RQRPA) \cite{16}. 
The $2\nu\beta\beta$-decay matrix elements calculated via pn-RQRPA and
full-RQRPA have been found significantly less sensitive to the
increasing strength of the particle-particle interaction 
\cite{15}-\cite{17}. This fact  
indicated that both the Pauli exclusion principle and 
proton-neutron pairing play an important role in the evaluation
of the many-body Green functions for the double beta decay.
In the meanwhile, some critical studies have shown that the renormalized 
QRPA has a few shortcomings, e.g., the violation of the Ikeda sum rule
\cite{17,18}. 
J. Hirsch et al. \cite{18,19} and J. Engel et al. \cite{20} studied the
validity of the renormalized QRPA within  schematic exactly solvable
models. Their studies confirmed that the renormalized QRPA offers 
advantages over the QRPA. But, they found some discrepancies 
between the exact and the RQRPA solutions after the point of collapse of 
the QRPA. However it is not clear whether their reults would  also hold 
for realistic calculations with  large model spaces and realistic
effective NN-interactions. In fact we do not know of any exactly solvable 
realistic model. The exact solution 
for the intermediate nuclear state discussed in Refs. \cite{18,19}
within a schematic one level model space, after the point of collapse of
QRPA, is below the initial ground state. This is not however the case of 
nuclei which undergo double beta decay . 
We note also that the violation of the Ikeda 
sum rule within the full-RQRPA with a large model space and a 
realistic effective NN-interaction  is rather small (about 10-20 $\%$)
\cite{21}.

We believe that the full-RQRPA is the most reliable method to deduce
the desired interesting lepton number non-conserving parameters from
the experimental lower limits on the half-lifes of neutrinoless 
double beta decay of heavier nuclei at present. To our knowledge
the full-RQRPA has been applied for the first time 
by \v Simkovic et al. \cite{21} 
in  calculations of  $0\nu\beta\beta$-decay. It has been found there 
by calculating the $0\nu\beta\beta$-decay of $^{76}Ge$ via
the pn-QRPA and full-RQRPA that by increasing the model space 
the pn-QRPA results are extremely sensitive to the renormalization
of the particle-particle interaction in the physically acceptable
region of the nuclear strength. This behaviour was similar to the one
known from  $2\nu\beta\beta$-decay calculations. On the other hand 
the full-RQRPA results show an increased stability in respect
to the stregth of the particle - particle interaction
with increasing model space.  This then sugests that the
quenching of the $2\nu\beta\beta$-decay matrix elements by
the pn-QRPA is not a special phenomenon for the $2\nu\beta\beta$-decay process
but is a common feature to all many-body Green functions defining
second order processes. As this quenching seems to have
its origin in an inaccuracy of the calculation steming from
the quasiboson approximation scheme it is necessary to recalculate
the  $0\nu\beta\beta$-decay matrix elements with the full-RQRPA,
which takes into account the Pauli exclusion principle and therefore
one can expect to deduce
more reliable limits on the lepton number non-conserving
parameters .

In this article, we shall study the nuclear matrix elements of
light neutrino mass mediated mode of $0\nu\beta\beta$-decay
of $^{76}Ge$, $^{100}Mo$, $^{116}Cd$, $^{128}Te$ and $^{136}Xe$.
The most stringent experimental lower limit of the 
life-time and large phase-space factors favour especially
these processes for extracting an upper limit on the effective neutrino 
mass. Our first consideration is to find out whether the strong sensitivity
of the $0\nu\beta\beta$-decay matrix elements within the pn-QRPA is 
a general feature to all double beta decay transitions. 
For that purpose we perform our calculations with a considerably 
large Hilbert model space . 
Our second task is to calculate the
$0\nu\beta\beta$-decay matrix elements  within the full-RQRPA 
and investigate in this way whether one might be able
to extract more accurate values on the 
effective neutrino mass.

\section{Theory}

In the case of  the light neutrino mass mediated mode of the 
$0\nu\beta\beta$-decay the weak beta decay Hamiltonian acquires the form:
\begin{equation}
{\cal H}^\beta(x)=\frac{G_{\text{F}}}{\sqrt{2}} 2
\left[\bar{e}_{\text{L}}(x)\gamma_\alpha \nu_{\text{e L}}(x)\right]
j_\alpha(x) + \text{h.c.},
\label{eq:2}   
\end{equation}
where $j_{\alpha}(x)$ is the strangeness conserving charged 
hadron currents  and
$e_{\text{L}}(x)$ and $\nu_{\text{e L}}(x)$ 
are operators of the left
components of fields of the electron and electron neutrino, respectively. 
We suppose that  neutrino mixing does take place according to,
\begin{equation}
\nu_{eL}=\sum_k ~U^L_{ek}~\chi_{kL},
\label{eq:3}   
\end{equation}
where, $\chi_{kL}$ is the field of the Majorana neutrinos with mass
$m_k$ and $U^L_{ek}$ is unitary mixing matrix. 

If we consider the usual approximations, i.e., non-relativistic momentum
approximation for hadron currents and long-wave approximation, and replace  
the energies of the outgoing electrons in the denominators with the half
of the available energy for this process, we get the following
matrix element: 
\begin{eqnarray}
<{f}|S^{{(2)}}|{i}>
&=& \frac{i}{2(2\pi)^3}
\left(\frac{G_{{F}}}{~\sqrt{2}}\right)^{{2}}
\frac{1}{\sqrt{p_{10}p_{20}}}~ <m_\nu>\frac{g^2_A}{R} \times
\nonumber \\
&& \bar{u}(p_{{1}})(1-\gamma_{\text{5}})C{\bar{u}}^T(p_{{2}})
\label{eq:4}   
\end{eqnarray}
Here, $p_{{1}}$, $p_{{2}}$ are the four-vector momenta of the electrons 
and $E^i$, $E^f$ are respectively the energies of the initial and final 
nuclear states.  
The effective neutrino mass is given as follows:
 \begin{equation}
<m_\nu> = \sum_{{j}} |U^{}_{{ej}}|^2 
m_{{j}} e^{ {{i}}\alpha_{{j}} }, 
\label{eq:5}   
\end{equation}
where $\exp({i}\alpha_{{j}})$ is the CP eigenvalue of the neutrino
mass eigenstate $|\chi_{{j}}>$. The nuclear matrix element of the process
takes the form: 
\begin{equation}
M^{0\nu}_{mass}=M^{0\nu}_{GT}-\left(\frac{g_V}{g_A}\right)^2M_F^{0\nu},
\label{eq:6}   
\end{equation}
The Gamow-Teller and Fermi nuclear matrix elements take the form
\begin{eqnarray}
{{ {M^{0\nu}_{{GT}}}}}=R\sum_{{mJ},L}
\int^{\infty}_{0}q^2 dq 
\frac{<0^+_f\parallel {\cal O}^{0\nu}_{GT}(qr;L,J){\parallel J m>}
{<J m\parallel} {\cal O}^{0\nu}_{GT}(qr;L,J)\parallel 0^+_{i}>}
{q_0 ~~[~ q_0 ~+ ~{E_J^m} ~- ~( E^i~ +~ E^f )/2~ ]},
\label{eq:7}   
\end{eqnarray}
 \begin{eqnarray}
{
{{M^{0\nu}_{{F}}}}=R\sum_{{mJ}}
\int^{\infty}_{0}q^2 dq
\frac{<0^+_f\parallel { O}^{0\nu}_{F}(qr,J){\parallel J m>}
{<J m\parallel} {O}^{0\nu}_{F}(qr,J)\parallel 0^+_{i}>}
{q_0 ~~[~ q_0 ~+ ~{E_J^m} ~- ~( E^i~ +~ E^f )/2~ ]},}
\label{eq:8}   
\end{eqnarray}
 with
\begin{equation}
{{\cal O}^{0\nu}_{GT}(qr,J)}=\sum_k \tau^+_k~ 
2\sqrt{2}~i^L~j_L(qr_k)~\{Y_L\otimes\sigma(k)_1\}_J,
\label{eq:9}   
\end{equation}
\begin{equation}
{{\cal O}^{0\nu}_{F}(qr,J)}= \sum_k \tau^+_k 
2\sqrt{2}~i^L~j_L(qr_k)~Y_J,
\label{eq:10}   
\end{equation}
Here, $R=r_0A^{1/3}$ is the nuclear radius ($r_0=1.1 $ fm),
$g_V ~=$ 1.00,  $g_A ~=$ 1.25 and $q_0 \approx q$ for light neutrino. 
In the formulae of the 
$M^{0\nu}_{GT}$ and $M^{0\nu}_{F}$ in Eqs.\ 
(\ref{eq:7}) and (\ref{eq:8})  
it is somehow difficult to include the correlation function
of the two interacting nucleons. Some attempts  
have been  made by Krmpoti\'c and Sharma \cite{11}. However, 
the operators  $M^{0\nu}_{GT}$ and $M^{0\nu}_{F}$ 
in (\ref{eq:7}) and (\ref{eq:8})    
 containing two one-body matrix elements 
are usually transformed to ones containing two-body matrix elements
in relative coordinates by using the second quantization formalism.
We then obtain:
\begin{equation}
 \left .
 \begin{array}{c}
  M_{GT}^{0\nu} \\
  M^{0\nu}_{F}
 \end{array}
 \right\}
= \left\langle H(r_{12}) 
\begin{array}{c}
{\bf \sigma}_1 \cdot {\bf\sigma}_2 \\
1
\end{array}
\right\rangle,
\label{eq:11}   
\end{equation}
\begin{eqnarray}
<O_{12}>&=&
\sum_{{k l \acute{k} \acute{l} } \atop {J^{\pi}
m_i m_f {\cal J}  }}
~(-)^{j_{l}+j_{k'}+J+{\cal J}}(2{\cal J}+1)
\left\{
\begin{array}{ccc}
j_k &j_l &J\\
j_{l'}&j_{k'}&{\cal J}
\end{array}
\right\}\nonumber \\
&&~~~~~~~~~~~
\times<pk,pk';{\cal J}|f(r_{12})\tau_1^+ \tau_2^+ {\cal O}_{12}
f(r_{12})|nl,nl';{\cal J}>
\nonumber \\
&&\times < 0_f^+ \parallel 
\widetilde{[c^+_{pk'}{\tilde{c}}_{nl'}]_J} \parallel J^\pi m_f>
<J^\pi m_f|J^\pi m_i>
<J^\pi m_i \parallel [c^+_{pk}{\tilde{c}}_{nl}]_J \parallel 
0^+_i >.
\label{eq:12}   
\end{eqnarray}
The short-range correlations
between the two interacting nucleons are now taken into account by
a correlation function \cite{5,10}. 
\begin{equation}
f(r_{12})=1-e^{-a r_{12}^2}(1-b r^2_{12})
\label{eq:13}   
\end{equation}
with $a = 1.1 fm^{-2}$, $b = 0.68 fm^{-2}$. 
The neutrino-potential $H(r_{12})$  takes the form
\begin{equation}
H(r)=\frac{2}{\pi}
\frac{R}{r} \int_{0}^{\infty} 
\frac{\sin(qr)}{q+(\Omega^{m_i}_{J^\pi}+\Omega^{m_f}_{J^\pi})/2}
\frac{1}{(1+q^2/{\Lambda}^2)^4}
d{q}.
\label{eq:14}   
\end{equation}
The parameter $\Lambda $ of the dipole shape nucleon form factor is
chosen to be 0.85 GeV \cite{5,10}.  
$\Omega^{m_i}_{J^\pi} = E^{m_i}_{J^\pi}-E^{i}_{0^+}$ and
$\Omega^{m_f}_{J^\pi} = E^{m_f}_{J^\pi}-E^{f}_{0^+}$.

For the half-life of the $0\nu\beta\beta$-decay we obtain:
\begin{equation}
[T_{1/2}^{0\nu}]^{-1} = G_{01} (M^{0\nu}_{mass})^2 
\left(\frac{<m_\nu>}{m_e}\right)^2. 
\label{eq:15}  
\end{equation}
$G_{01}$ is the integrated kinematical factor for the
$0^+_i\rightarrow 0^+_f$ transition \cite{1,3,23}.

In order to calculate the nuclear matrix element $M^{0\nu}_{mass}$
the full set of the intermediate nuclear states has to be constructed
e.g., by the QRPA or RQRPA diagonalization.
The full-RQRPA, which describes the excited states of the even-even
nucleus, has been studied in Ref. \cite{16} and the pn-QRPA, which is a
special case of the full-RQRPA  in Refs. \cite{12,13,14}.
Therefore, here we shall present only the formulae relevant to this work.  

The quasiparticle creation and annihilation operators 
($a^{+}_{\mu a m_{a}}$ and $a^{}_{\mu a m_{a}}$, $\mu =1,2$) 
are defined through the 
Hartree- Fock- Bogoliubov (HFB) transformation,
 which includes proton-proton, neutron-neutron
and proton-neutron pairing  \cite{22,23}:
\begin{equation}  \left( \matrix{ c^{+}_{p k m_{k} } \cr
c^{+}_{n k m_{k}} \cr {\tilde{c}}_{p k m_{k} } \cr 
{\tilde{c}}_{n k {\tilde{m}}_{k}} 
}\right) = \left( \matrix{ 
u_{k 1 p} & u_{k 2 p} & -v_{k 1 p} & -v_{k 2 p} \cr 
u_{k 1 n} & u_{k 2 n} & -v_{k 1 n} & -v_{k 2 n} \cr
v_{k 1 p} & v_{k 2 p} & u_{k 1 p} & u_{k 2 p} \cr  
v_{k 1 n} & v_{k 2 n} & u_{k 1 n} & u_{k 2 n} }\right)
\left( \matrix{ a^{+}_{1 k m_{k}} \cr
a^{+}_{2 k m_{k}} \cr {\tilde{a}}_{1 k m_{k}} \cr 
{\tilde{a}}_{2 k m_{k}} }\right).
\label{eq:16}  
\end{equation} 
Here, $c^{+}_{\tau a m_{a}}$ ($c^{}_{\tau a m_{a}}$) denotes
the particle creation (annihilation) operator acting on
a single particle level  k with isospin $\tau = p,n$. 
The tilde  $\sim$ indicates the time reversed states
${\tilde{c}}_{\tau k m_k}=(-1)^{j_k-m_k}c_{\tau k -m_k}$ etc.

In the full-RQRPA the commutator
of two bifermion operators is replaced with its expectation value in
the correlated QRPA ground state $|0^+_{QRPA}>$
(renormalized quasiboson approximation). We have
\begin{eqnarray}
&&
\big<0^+_{QRPA}\big|\big
[A^{}_{\mu \nu}(k, l, J, M),A^+_{\mu' \nu'}(k', l', J, M)\big]
\big|0^+_{QRPA}\big> 
\nonumber \\ 
&&=n(k\mu, l\nu) n(k'\mu', l'\nu') 
\Big( \delta_{kk'}\delta_{\mu \mu' }\delta_{ll'}
\delta_{\nu\nu'} -
\delta_{lk'}\delta_{\nu \mu'}\delta_{kl'}
\delta_{\mu \nu'}(-1)^{j_{k}+j_{l}-J}\Big) \nonumber \\
&&\times \underbrace{
\Big\{1
\,-\,\frac{1}{\hat{\jmath}_{l}}
<0^+_{QRPA}|[a^+_{\nu l}{\tilde{a}}_{\nu l}]_{00}|0^+_{QRPA}>
\,-\,\frac{1}{\hat{\jmath}_{k}}
<0^+_{QRPA}|[a^+_{\mu k}{\tilde{a}}_{\mu \tilde{k}}]_{00}|0^+_{QRPA}>
\Big\}
}_{
=:\displaystyle {\cal D}_{\mu k, \nu  l; J^\pi}
},
\label{eq:17}  
\end{eqnarray}
with  $\hat{\jmath}_k=\sqrt{2j_k+1}$.
 The operator 
$A^+_{\mu \nu}(k,l,J,M)$  creates 
a pair of quasiparticles coupled to angular momentum J with projection M.
\begin{eqnarray}
A^+_{\mu \nu}(k, l, J, M) &=& n(k\mu, l\nu) \sum^{}_{m_k , m_l }
C^{J M}_{j_k m_k j_l m_l } a^+_{\mu k m_k} a^+_{\nu l m_l},
\nonumber \\
n(k\mu, l \nu)&=&
(1+(-1)^J\delta_{kl}\delta_{\mu \nu})/(1+\delta_{kl}
\delta_{\mu \nu})^{3/2}.
\label{eq:18}  
\end{eqnarray}
If we replace $|0^+_{QRPA}>$ in Eq. (\ref{eq:17}) with the uncorrelated
HFB ground state, we obtain the quasiboson approximation (i.e. 
${\cal D}_{\mu k, \nu  l; J^\pi}=1$), which violates the Pauli 
exclusion principle by neglecting the terms coming from the 
commutator.  The full-RQRPA takes into account the Pauli exclusion
principle more carefully. The coefficients 
${\cal D}_{\mu k, \nu  l; J^\pi}$, which renormalize the particle-hole
and particle-particle interaction entering the 
${\cal A}$ and ${\cal B}$ matrices of the full-RQRPA equation 
\begin{equation}
\left(\begin{array}{cc}
 {\cal A}& {\cal B}\\
 {\cal B}& {\cal A}\end{array}\right)_{J^\pi}
\left(
\begin{array}{c} {\overline{X}}^m\\ {\overline{Y}}^m \end{array}
\right)_{J^\pi}
= \Omega^m_{J^\pi}
\left(
\begin{array}{cc}
1&0\\
0&-1
\end{array}
\right)
\left(
\begin{array}{c}
{\overline{X}}^m\\
{\overline{Y}}^m
\end{array}
\right)_{J^\pi},
\label{eq:19}  
\end{equation}
are determined by solving numerically the system of non-linear 
equations \cite{15,16}:
\begin{eqnarray}
{\cal D}_{k\mu l\nu J^\pi} &=& 1-\frac{1}{\hat{\jmath}_k^2} 
\sum_{k'\mu' \atop J'^{\pi'} m}
{\cal D}_{k\mu k'\mu'J'^{\pi'}}\hat{J}'^2 \big|
{\overline{Y}}^m_{\mu \mu'}(k, k', J'^{\pi'})  \big|^2
\nonumber \\
&& ~~~~~-\frac{1}{\hat{\jmath}_l^2}\sum_{l'\nu' \atop J'^{\pi'} m}
{\cal D}_{l\nu
l'\nu'J'^{\pi'}}\hat{J}'^2 \big |{\overline{Y}}^m_{\nu \nu'}(l, l', J'^{\pi'})
  \big|^2 .
\label{eq:20}  
\end{eqnarray}
The selfconsistent scheme of the calculation of forward-
(backward-) going free variational amplitude   
${\overline{X}}^m_{}$ (${\overline{Y}}^m_{}$), 
energies of the excited states 
$\Omega^m_{J^\pi}$ and coefficients ${\cal D}_{\mu k, \nu  l; J^\pi}$ 
is a double iterative problem which requires the solution of 
coupled-non-linear equations. We note that in the limit 
${\cal D}_{\mu k, \nu  l; J^\pi}=1$ and 
in the case  proton-neutron pairing is switched off, 
the solution of the full-RQRPA coincides with the solution 
of the pn-QRPA \cite{13}.

For the one-body densities in the full-RQRPA we can write:
\begin{eqnarray}
<J^\pi m_i \parallel [c^+_{pk}{\tilde{c}}_{nl}]_J \parallel 0^+_i>&=&
\sqrt{2J+1}\sum_{\mu,\nu = 1 , 2} m(\mu k,\nu l) 
\left [
u_{k \mu p}^{(i)} v_{l \nu n}^{(i)} {\overline{X}}^{m_i}_{\mu\nu}(k,l,J^\pi)
\right.\nonumber \\ && \left.
+v_{k \mu p}^{(i)} u_{l \nu n}^{(i)} {\overline{Y}}^{m_i}_{\mu\nu}(k,l,J^\pi)
\right ]
\sqrt{{\cal D}^{(i)}_{k \mu l \nu l J^\pi}},
\label{eq:21}   
\end{eqnarray}
\begin{eqnarray}
<0_f^+ \parallel \widetilde{ [c^+_{pk'}{\tilde{c}}_{nl'}]_J} 
\parallel J^\pi m_f>&=&
\sqrt{2J+1}\sum_{\mu,\nu=1,2}m(\mu k',\nu l') 
\left [
v_{k' \mu p}^{(f)} u_{l' \nu n}^{(f)} 
{\overline{X}}^{m_f}_{\mu\nu}(k',l',J^\pi)
\right.\nonumber \\ && \left.
+u_{k' \mu p}^{(f)} v_{l' \nu n}^{(f)} 
{\overline{Y}}^{m_f}_{\mu\nu}(k',l',J^\pi)
\right ]
\sqrt{{\cal D}^{(f)}_{k' \mu l' \nu J^\pi}},
\label{eq:22}   
\end{eqnarray}
with 
$m(\mu a,\nu b)=\frac{1+(-1)^J\delta_{\mu \nu}\delta_{ab}}{(1+\delta_{\mu\nu}
\delta_{ab})^{1/2}}$. We note that the  
${\overline{X}}^{m}_{\mu\nu}(k,l,J^\pi)$ and
${\overline{Y}}^{m}_{\mu\nu}(k,l,J^\pi)$ amplitudes are calculated 
by the renormalized QRPA equation only for the configurations 
$\mu a \leq \nu b$ (i.e., 
$\mu = \nu$ and the orbitals are ordered $a \leq b$ and 
$\mu = 1$, $\nu = 2$ and the orbitals are not ordered) \cite{23}.
For different configurations 
${\overline{X}}^{m}_{\mu\nu}(k,l,J^\pi)$ and
${\overline{Y}}^{m}_{\mu\nu}(k,l,J^\pi)$ in Eqs. 
(\ref{eq:21}) and (\ref{eq:22}) are given as follows:
\begin{equation}
{\overline{X}}^{{m}}_{\mu \nu}(k,l,J^\pi)
 = -(-1)^{{{j}}_{{k}}+{{j}}_{{l}}-{J}}
{\overline{X}}^{{m}}_{\nu \mu}(l,k,J^\pi), 
\label{eq:23}    
\end{equation} 
\begin{equation}
{\overline{Y}}^{{m}}_{\mu \nu}(k,l,J^\pi)
 = -(-1)^{{{j}}_{{k}}+{{j}}_{{l}}-{J}}
{\overline{Y}}^{{m}}_{\nu \mu}(l,k,J^\pi). 
\label{eq:24}    
\end{equation} 
The index i (f) indicates that the quasiparticles and the excited
states of the nucleus are defined with respect to the initial (final)
nuclear ground state $|0^+_i>$ ($|0^+_f>$). 
We note that for ${\cal D}_{k \mu l \nu J^\pi}=1$ 
(i.e. there is no renormalization) and $u_{\text{2p}} =
\upsilon_{\text{2p}} = u_{\text{1n}} = \upsilon_{\text{1n}} = 0$ (i.e.
there is no proton-neutron pairing), Eqs.\ (\ref{eq:21}) and
(\ref{eq:22}) reduce to the expressions of the pn-QRPA \cite{7}-\cite{9}.
The overlap between two
intermediate nuclear states belonging to two different sets is given by:
\begin{equation}
<J^\pi m_f|J^\pi m_i>=\sum_{\mu k \leq \nu l}
\left [
{\overline{X}}^{m_f}_{\mu\nu}(kl,J^\pi)
{\overline{X}}^{m_i}_{\mu\nu}(kl,J^\pi)
-{\overline{Y}}^{m_f}_{\mu\nu}(kl,J^\pi)
{\overline{Y}}^{m_i}_{\mu\nu}(kl,J^\pi)
\right ].
\label{eq:25}   
\end{equation}

\section{Calculation and discussion}

We applied  both the pn-QRPA and the full-RQRPA methods to calculate
the $0\nu\beta\beta$-decay of the $A = 76, 100, 116, 128, 136$ systems.
In our calculations we tried to use as large as possible Hilbert model
spaces limited only  by the power of the available  computers.  
We have taken the folowing 
single particle model spaces for both, protons and neutrons :\\
(i) For $^{76}Ge \rightarrow ^{76}Se$, $^{100}Mo \rightarrow ^{100}Ru$ 
and $^{116}Cd \rightarrow ^{116}Sn$ 
the model space comprises 21 levels:
$0s^{}_{1/2}$, $0p^{}_{1/2}$, $0p^{}_{3/2}$,  $1s^{}_{1/2}$, 
$0d^{}_{3/2}$, $0d^{}_{5/2}$, $1p^{}_{1/2}$, $1p^{}_{3/2}$,
 $0f^{}_{5/2}$, $0f^{}_{7/2}$, $2s^{}_{1/2}$, $1d^{}_{3/2}$, $1d^{}_{5/2}$,
 $0g^{}_{7/2}$, $0g^{}_{9/2}$, $2p^{}_{1/2}$, $2p^{}_{3/2}$,
 $1f^{}_{5/2}$, $1f^{}_{7/2}$, $0h^{}_{9/2}$, $0h^{}_{11/2}$.\\
(ii) For $^{128}Te \rightarrow ^{128}Xe$ and 
$^{136}Xe \rightarrow ^{136}Ru$ we used 20 levels:
$1s^{}_{1/2}$, 
$0d^{}_{3/2}$, $0d^{}_{5/2}$, $1p^{}_{1/2}$, $1p^{}_{3/2}$,
 $0f^{}_{5/2}$, $0f^{}_{7/2}$, $2s^{}_{1/2}$, $1d^{}_{3/2}$, $1d^{}_{5/2}$,
 $0g^{}_{7/2}$, $0g^{}_{9/2}$, $2p^{}_{1/2}$, $2p^{}_{3/2}$,
 $1f^{}_{5/2}$, $1f^{}_{7/2}$, $0h^{}_{9/2}$, $0h^{}_{11/2}$,
  $0i^{}_{11/2}$, $0i^{}_{13/2}$. \\
These model spaces are considerably larger as those used in any previous 
pn-QRPA calculations \cite{7}-\cite{14}. 
The single particle energies have been calculated with a 
Coulomb-corrected Woods-Saxon potential. The nucleon-nucleon interaction
used to calculate the nuclear wave functions is based on the 
Brueckner G-matrix derived from the Bonn one-boson-exchange potential,
which in principle is a more consistent and better description of the
NN-interaction in nuclei. The Brueckner reaction matrix is obtained
by solving the Bethe-Goldstone equation \cite{24}. The pn-QRPA is based
on a Bardeen-Cooper-Schrieffer (BCS) transformation including 
proton-proton and neutron-neutron pairing correlations. In the case of
the full-RQRPA the single quasiparticle energies and 
occupation amplitudes have been found by solving the HFB equation with
p-n pairing in the above mentioned space \cite{22}. 
Since our model spaces are finite, all pairing potential 
gaps are renormalized to the empirical gaps by the strength parameters
$d_{pp}$, $d_{nn}$ a $d_{pn}$. Technically this is achieved by performing
a BCS and HFB calculation and comparing the obtained pairing gaps with
the ones extracted from the empirical separation energies in a manner
described in Ref. \cite{22}. 
The renormalization parameters $d_{pp}$, $d_{nn}$ and $d_{pn}$
together with the experimental proton ( $\Delta ^{exp}_{p}$ ), neutron 
( $\Delta ^{exp}_{n}$ ) and proton - neutron ( $\delta
^{exp}_{pn}$ ) pairing gaps, for all studied nuclear systems, are listed
in Table \ref{table1}. The experimental pairing gaps are defined by
Moeller and Nix \cite{25,26}. By glancing at the Table \ref{table1} we see 
that the $d_{pp}$ and $d_{nn}$ values are close to unity and $d_{pn}$ is 
higher than these values. It is because for spherical nuclei the 
HFB-transformation can only describe correlations for pairs with
J=0 and T=1 and not for pairs with J=0 and T=0. The T=0 
pairing is effectively taken into account by the renormalization 
of the T=1 J=0 n-p interaction leading to a higher value of $d_{pn}$. 
However, we do not want to focuse our attention to the problem of 
proton-neutron pairing, which seems to play a significant role 
in the QRPA calculation of the double beta decay process \cite{16,23} 
but it is less important in the case of the renormalized QRPA calculations
\cite{21}. The proton-neutron pairing problem is extensively discussed in 
\cite{22,27} and references cited there. 
We note that in the BCS and HFB calculations we neglected
the mixing of different "n" but the same "ljm" orbitals.
We suppose that shell mixing is not significantly affecting 
the BCS and HFB solutions because their off-diagonal 
pairing matrix elements are quite small. 

After settling the values of the pairing parameters, the parameters which 
remain to be fixed are the particle- particle and the particle- 
hole strengths. 
The particle - particle and particle - hole channels of the G-matrix 
interaction are renormalized 
by introducing the two parameters $g^{}_{pp}$ and $g^{}_{ph}$,
which, in principle, should be close to unity. Our adopted values were 
$g_{\text{ph}} = 0.8$, as in our previous calculations \cite{21,23}
and  $g_{pp}$ is varied in the interval
0.70 - 1.30 which can be regarded as physical.

The nuclear matrix elements $M^{0\nu}_{mass}$ for the most interesting 
nuclei obtained within the
pn-QRPA are shown in Fig. 1 (a). We see that $M^{0\nu}_{mass}$ for 
the A=76, 100, 116, 128 and 136 systems, becomes singular with 
increasing 
strength of the particle-particle interaction and even  crosses zero
in the physically acceptable region of the parameter $g_{pp}$ . 
This behavior has not been  found in the previous calculations 
\cite{7}-\cite{10} because the model spaces used there, were too small. 
To our knowledge
only Krmpoti\'c and Sharma \cite{11} have studied the model space 
dependence of the  $0\nu\beta\beta$-decay matrix elements.
For a relatively large model space 
they found for the $0\nu\beta\beta$-decay of $^{48}Ca$ a similar
behavior. However this was further out of the physicall region
of the strength renormalization
parameters. We note that there is a principal difference between
our calculations and those of Krmpoti\'c and Sharma. They used a zero 
range $\delta$-force interaction 
and a different treatment to the two-nucleon correlation function, which 
were incorporated in the formalism in Eqs.\ (\ref{eq:6}-\ref{eq:10}).
To our opinion their type of two-nucleon correlations influences the
result unsignificantly. The two-nucleon correlations  presented in this
work give an effect of about 30-40$\%$. We should note further that 
$^{48}Ca$ is a closed shell nucleus and therefore it is not very suitable
for the QRPA calculation.

The QRPA quenching mechanism for $M^{0\nu}_{mass}$ in Fig. 1 (a) 
has its origin, exactly in the same way as 
the quenching mechanism of $2\nu\beta\beta$-decay matrix 
elements, i.e., in the 
generation of too much ground state correlations with increasing $g_{pp}$ 
near the collapse of the QRPA. A larger model space means more 
ground state correlations, i.e., a collapse of the QRPA solution
for smaller $g_{pp}$. 
As a consequence the validity of the
quasiboson approximation in the evaluation of the $0\nu\beta\beta$-decay
matrix elements is questioned because of the generation of too much ground
state correlations. For that reason it is necessary to perform 
the calculation of $M^{0\nu}_{mass}$ in the framework of the 
renormalized QRPA, which takes into account the Pauli exclusion 
principle in an approximate way and also considers 
proton-neutron pairing correlations. 
 In Fig. 1 (b) we present our results 
with the full-RQRPA method. From the comparison with the Fig. 1 (a)
it follows that the inclusion of ground state correlations beyond
the QRPA in the calculation of $M^{0\nu}_{mass}$ removes the
difficulties associated with the extreme sensitivity of
$M^{0\nu}_{mass}$ on the particle-particle strength. 
The strong differences between the results of both methods indicate that the
Pauli exclusion principle plays an important role in the evaluation
of the $0\nu\beta\beta$-decay. 

A weakly dependence of the $M^{0\nu}_{mass}$ on $g_{pp}$ 
inside the physical range $0.8
\le g_{pp} \le 1.2$ allows us to have
more confidence for deducing the effective neutrino mass $<m_\nu>$ from the 
available experimental lower limits  on the half-lives of
the $0\nu\beta\beta$-decays $T^{0\nu -exp}_{1/2}$.  
The nuclear matrix elements $M^{0\nu}_{mass}$ obtained 
within the full-RQRPA for $g_{pp}=1.0$, the integrated kinematical
factors $G_{01}$, $T^{0\nu -exp}_{1/2}$ 
and the limits on the effective neutrino mass
$|<m_{\nu}>|$ deduced from $M^{0\nu}_{mass}$ ($g_{pp}=1.0$) and 
$T^{0\nu -exp}_{1/2}$ are listed in Table \ref{table2}.
A more stringent upper limit on $<m_\nu>$ is favored by large values of 
$M^{0\nu}_{mass}$, $G_{01}$ and  $T^{0\nu -exp}_{1/2}$. We see that  
the most stringent upper limit $<m_\nu> ~ \le ~1.1$ eV
is deduced for the A=76  system mainly because of the unbelievable
high upper limit on $T^{0\nu -exp}_{1/2}$ given by the Heidelberg- Moscow
collaboration \cite{28}. Also of interest is the value 
$T^{0\nu {-1eV}}_{1/2}$ (see Table \ref{table2}), which
is calculated by assuming $|<m_{\nu}>|$ ={ 1eV}.  This value indicates 
that for further experimental measurements the most perspective
candidate is $^{100}Mo$.

\section{Summary and conclusion}

In summary, we have studied the nuclear matrix elements entering
the light neutrino mediated mode of the neutrinoless double beta decay
of some experimentally  interesting nuclear systems,
A=76, 100, 116, 128 and 136 . The calculations
have been performed within both, the pn-QRPA and the full-RQRPA  methods
with a large Hilbert model space. We have found 
that in the framework of the pn-QRPA for a large enough model space
the $0\nu\beta\beta$-decay matrix elements demonstrate an
instability with respect to the renormalization 
of the particle-particle strength, similar to the one known from 
the $2\nu\beta\beta$-decay mode. The value of the matrix
element crosses zero and it is then difficult to  make definite
rate predictions. We believe that
the common quenching phenomenon which is independent
of the studied nucleus and double beta decay process could have its origin
only in the approximation scheme. 
The full-RQRPA which includes proton-neutron pairing  and 
the Pauli effect of fermion pairs goes
beyond the quasi-boson approximation. The inclusion of the Pauli principle
eliminates the instabilities that plaque the pn-QRPA.
The $0\nu\beta\beta$-decay matrix elements calculated via the full-RQRPA
are stable in respect to the changes of the particle-particle force 
 and it allows us to deduce more accurate limits on the effective neutrino
mass. The largest $0\nu\beta\beta$-decay matrix elements 
are 4.22 and 3.28 associated with the A = 100 and 128 systems, respectively.
A large value of the matrix element together with a large value
of the kinematical factor favour especially $^{100}Mo$ 
for further experimental study of the $0\nu\beta\beta$-decay.
At present the A= 76 and 128 systems provide us with the most 
stringent limit on the effective neutrino  mass i.e., 1.1 - 1.2 eV.

\newpage

\widetext
\begin{table}[t]
\caption{Experimental proton ( $\Delta ^{exp}_{p}$ ), neutron 
( $\Delta ^{exp}_{n}$ ) and proton - neutron ( $\delta
^{exp}_{pn}$ ) pairing gaps and
renormalization constants of the proton - proton (
$d_{pp}$ ),
neutron - neutron ( $d_{nn}$ ) and proton - neutron ( $d_{pn}$ )
pairing interactions for all nuclei studied here.}
\label{table1}
\begin{tabular}{ccccc} 
Nucleus ~ ( ${\Delta }^{exp}_{p}$, ${\Delta }^{exp}_{n}$, ${\delta }
^{exp}_{pn}$ ) & model & $d_{pp}$ & $d_{nn}$ & $d_{pn}$ \\ 
~~~~~~ ( [MeV] ) & space & & & \\ \tableline
 $ ^{76}_{32}Ge_{44}$ ~ (1.561, 1.535, 0.388) 
 & 21 level & 0.899 & 1.028 & 1.506 \\ 
 $ ^{76}_{34}Se_{42}$ ~ (1.751, 1.710, 0.459) 
 & 21 level & 0.934 & 1.059 & 1.325 \\ 
 $ ^{100}_{42}Mo_{58}$ ~ (1.612, 1.358, 0.635) 
 & 21 level & 0.980 & 0.923 & 1.766 \\
 $ ^{100}_{44}Ru_{56}$ ~ (1.548, 1.296, 0.277) 
 & 21 level & 1.002 & 0.945 & 1.568 \\
 $ ^{116}_{48}Cd_{68}$ ~ (1.493, 1.377, 0.371) 
 & 21 level & 0.953 & 0.922 & 1.822 \\
 $ ^{116}_{50}Cd_{66}$ ~ (1.763, 1.204, 0.128) 
 & 21 level & 1.00 & 0.873 & 1.460 \\
 $ ^{128}_{52}Te_{76}$ ~ (1.127, 1.177, 0.149) 
 & 20 level & 0.873 & 0.942 & 1.780 \\
 $ ^{128}_{54}Xe_{74}$ ~ (1.307, 1.266, 0.199) 
 & 20 level & 0.920 & 0.972 & 1.530 \\
 $ ^{136}_{54}Xe_{82}$ ~ (0.971, 1.408, 0.0) 
 & 20 level & 0.771 & 0.803 & 0.0 \\
 $ ^{136}_{56}Ba_{80}$ ~ (1.245, 1.032, 0.165) 
 & 20 level & 0.875 & 0.899 & 1.716 \\
\end{tabular}
\end{table}

\begin{table}[h]
\caption{ 
The nuclear matrix elements $M^{0\nu}_{mass}$ (see Eqs. (6)
and (11-12))  obtained 
within the full-RQRPA for $g_{pp}=1.0$, the integrated kinematical
factors $G_{01}$, the limits on the effective neutrino mass
$|<m_{\nu}>|$ deduced from the experimental limit of the 
$0\nu\beta\beta$-decay lifetime $T^{0\nu -exp}_{1/2}$ for the
nuclei studied in this work.
$T^{0\nu {-1eV}}_{1/2}$ is the calculated
 $0\nu\beta\beta$-decay half-life times assuming 
$|<m_{\nu}>|$ = 1eV.  }
\label{table2}
\begin{tabular}{cccccc} 
nucleus &  $M^{0\nu}_{mass}$ & $G_{01}$ &  $T^{0\nu{-1eV}}_{1/2}$ &
 $T^{0\nu{-exp}}_{1/2}$ & $|<m_{\nu }>|$ \\
 & & [$10^{-14}$ $years^{-1}$] & [years] & [years] ref. & [eV]    \\ \hline
 & & & & & \\
$^{76}Ge$ &  1.86 & 0.7928 & $9.5\times{10^{24}}$ &
$\ge 7.4\times{10^{24}}$ (90\% C.L.) \cite{28} & $\le 1.1$ \\
$^{100}Mo$ & 4.22 & 5.731  & $2.6\times{10^{23}}$ &
$\ge 4.4\times{10^{22}}$ (68\% C.L.) \cite{29} & $\le 2.4$ \\
$^{116}Cd$ & 2.47 & 6.237  & $6.9\times{10^{23}}$ &
$\ge 2.9\times{10^{22}}$ (90\% C.L.) \cite{30} & $\le 4.9$ \\
$^{128}Te$ & 3.28 & 0.2207 & $1.1\times{10^{25}}$ &
$\ge 7.3\times{10^{24}}$ (68\% C.L.) \cite{31} &  $\le 1.2$ \\
$^{136}Xe$ & 0.96 & 5.914  &  $3.4\times{10^{23}}$ &
$\ge 6.4\times{10^{23}}$ (90\% C.L.) \cite{32} & $\le 3.7$ \\
\end{tabular}
\end{table}

\newpage

\begin{figure}[htb]
\hspace{1.5cm}
\vspace{0.5cm}
\caption{The calculated nuclear matrix element $M^{0\nu}_{mass}$
for the $0\nu\beta\beta$-decay of $^{76}Ge$,  $^{100}Mo$,  $^{116}Cd$,  
$^{128}Te$ and $^{136}Xe$  as a function of the
particle-particle interaction strength $g_{pp}$. In (a) 
$M^{0\nu}_{mass}$ has been calculated with the pn-QRPA method, in (b)
with the full-RQRPA . }
\label{figa}
\end{figure}

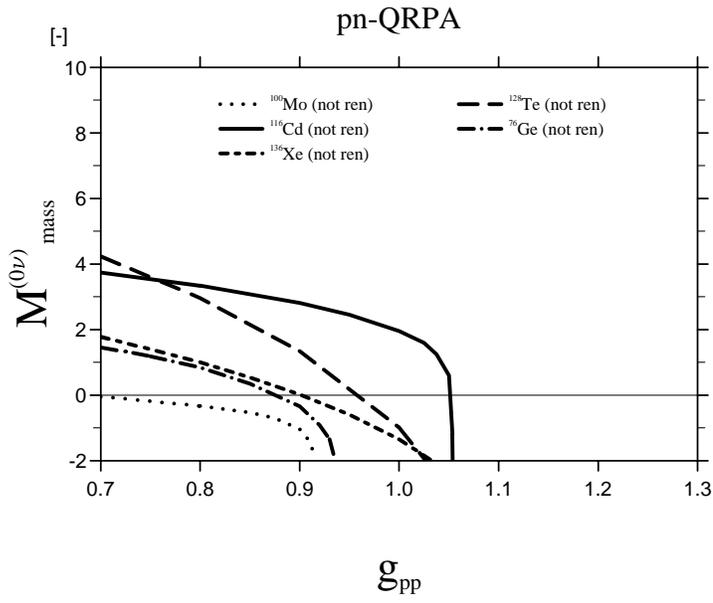
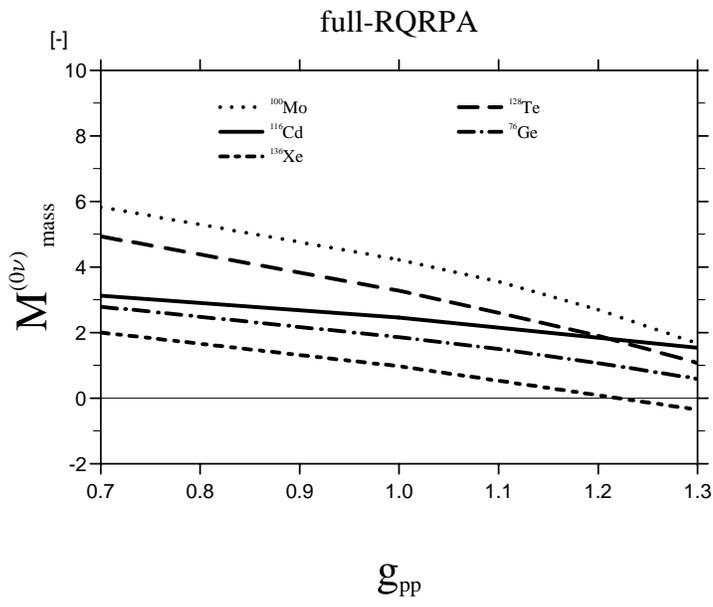
\end{document}